\documentstyle[preprint,aps,epsfig,axodraw]{revtex}

\begin{document}

\draft

\preprint{
\hfill$\vcenter{\hbox{\bf IFT-P.030/98} 
                \hbox{\bf MADPH-98-1058}
                \hbox{\bf UH-511-898-98} 
}$ }

\title{Signal and Backgrounds for the Single Production of \\
Scalar and Vector Leptoquarks at the LHC}
 
\author{J.\ E.\ Cieza Montalvo$^1$, O.\ J.\ P.\ \'Eboli$^{2,3}$, 
M.\ B.\ Magro$^3$, and P.\ G.\ Mercadante$^4$}

\address{$^1$Instituto de F\'{\i}sica, 
Universidade do Estado do Rio de Janeiro \\
 CEP $20559-900$ Rio de Janeiro, Brazil.}

\address{$^2$Instituto de F\'{\i}sica Te\'orica, Universidade Estadual 
Paulista\\
Rua Pamplona 145, S\~ao Paulo, SP 01405, Brazil.}

\address{$^3$Department of Physics, University of Wisconsin \\
Madison, WI 53706, USA.}

\address{$^4$Department of Physics $\&$ Astronomy, University of Hawaii\\ 
Honolulu, HI 96822, USA.}


\vskip -0.75cm 

\maketitle

\begin{abstract}
  
\vskip-5ex 

  We perform a detailed analyses of the potentiality of the CERN Large Hadron
  Collider to study the single production of leptoquarks via $ pp \to e^\pm q
  \to$ leptoquark $\to e^\pm q$, with $e^\pm$ generated by the splitting of
  photons radiated by the protons.  Working with the most general $SU(2)_L
  \otimes U(1)_Y$ invariant effective lagrangian for scalar and vector
  leptoquarks, we analyze in detail the leptoquark signals and backgrounds
  that lead to a final state containing an $e^\pm$ and a hard jet with
  approximately balanced transverse momenta.  Our results indicate that the
  LHC will be able to discover leptoquarks with masses up to 2--3 TeV,
  depending on their type, for Yukawa couplings of the order of the
  electromagnetic one.

\end{abstract}




\section{Introduction}

Leptoquarks, which are particles that carry simultaneously leptonic and
barionic numbers, provide a clear sign for many extensions of the Standard
Model (SM) that treat quarks and leptons in the same footing.  There are many
models exhibiting these new particles, such as technicolor \cite{tec},
composite models \cite{comp,af}, grand unified theories \cite{gut}, and
superstring-inspired models \cite{e6}. From the experimental point of view,
leptoquarks possess the striking signature of a peak in the invariant mass of
a charged lepton with a jet, which make their search much simpler without the
need of elaborate analyses of several final state topologies.

Pair production of leptoquarks in a hadronic collider takes place via
quark--quark and gluon--gluon fusions, being essentially model independent
since the leptoquark--gluon interaction is fixed by the $SU(3)_C$ gauge
invariance, with the only free parameter being an ``anomalous chromomagetic
moment'' for vector leptoquarks. On the other hand, single production is model
dependent because it takes place via leptoquark interactions with quarks and
leptons.  Notwithstanding, these two signals are complementary because they
allow us not only to reveal the existence of leptoquarks but also to determine
their properties such as mass and Yukawa couplings to quarks and leptons.

The direct search for leptoquarks with masses above a few hundred GeV can be
carried out only in the next generation of colliders. In fact, there have been
many studies of the production of leptoquarks in the future $pp$
\cite{fut:pp}, $ep$ \cite{buch,fut:ep}, $e^+e^-$ \cite{fut:ee}, $e^-e^-$
\cite{fut:elel}, $e\gamma$ \cite{fut:eg}, and $\gamma\gamma$ \cite{fut:gg}
colliders.  In particular, the usual studies for leptoquarks in hadronic
colliders concentrated on the processes \cite{thais}
\begin{eqnarray}
q + g~ &&\rightarrow \Phi_{\text{lq}} + \ell \; ,
\label{eq:sin}
\\
q + \bar{q}~ &&\rightarrow \Phi_{\text{lq}} + \bar{\Phi}_{\text{lq}} \; ,
\label{eq:qq}
\\
g + g~ &&\rightarrow \Phi_{\text{lq}} + \bar{\Phi}_{\text{lq}} \; ,
\label{eq:gg}
\end{eqnarray}
where $\ell = e^\pm$ ($\mu^\pm$) and we denoted scalar and vector leptoquarks
by $\Phi_{\text{lq}}$. These processes give rise to $e^+e^-$ pairs with large
transverse momenta accompanied by one or more jets.

In this work, we study the capability of the CERN Large Hadron Collider (LHC)
to unravel the existence of first generation leptoquarks through the final
state topology jet plus $e^\pm$.  This process was first analyzed in Ref.\ 
\cite{zerwas}, and it occurs via
\begin{equation}
p p \rightarrow e^\pm q \rightarrow \hbox{ leptoquark}
~\to e^\pm q \; ,
\end{equation}
where the $e^\pm$ originates from the splitting of a photon radiated by a
quark. This reaction leads an $e^\pm$--jet pair with balanced transverse
momenta, up to the detector resolution. This feature allows us to separate the
production mechanisms (\ref{eq:sin})--(\ref{eq:gg}) from the above reaction.
Therefore, this process provides one more handle to study the leptoquark
properties.

We performed a careful analyses of the signal and its respective backgrounds
for leptoquarks that couple to pairs $e^- u$, $e^+ u$, $e^- d$, or $e^+ d$,
assuming the most general effective Lagrangian that is invariant under
$SU(3)_C \otimes SU(2)_L \otimes U(1)_Y$ \cite{buch}.  We studied not only a
series of cuts to reduce the backgrounds, but also strategies to discriminate
among the several leptoquark types.  Our analysis improves the previous one
\cite{zerwas} since we considered all possible backgrounds as well as the most
general model for leptoquarks.

We show in this work that the single leptoquark search at the LHC can
discover, at the 3$\sigma$ level, leptoquarks with masses up to 2--3 TeV,
depending on their type, for an integrated luminosity of $100$ fb$^{-1}$, a
center-of-mass energy of $14$~TeV, and Yukawa couplings of the order of the
electromagnetic one. We also exhibit the region of the Yukawa coupling and
leptoquark mass plane that can be ruled out at the LHC.

The outline of this paper is as follows. In Sec.\ \ref{l:eff} we introduce the
$SU(2)_L \otimes U(1)_Y$ invariant effective Lagrangians that we analyzed and
present the available bounds on leptoquarks.  Sec.\ \ref{sign} contains a
detailed description of the leptoquark signal and its backgrounds, as well as
the cuts used to enhance the signal. We present our results in Sec.\ 
\ref{resu} and draw our conclusions in Sec.\ \ref{conc}.


\section{Models for leptoquark interactions}
\label{l:eff}

A natural hypothesis for theories beyond the SM is that they exhibit the gauge
symmetry $SU(2)_L \otimes U(1)_Y$ above the electroweak symmetry breaking
scale $v$. Therefore, we imposed this symmetry on the leptoquark interactions.
Moreover, in order to avoid strong bounds coming from the proton lifetime
experiments, we required baryon ($B$) and lepton ($L$) number conservation,
which forbids the leptoquarks to couple to pairs of quarks or leptons. The
most general effective Lagrangian for scalar and vector leptoquarks satisfying
the above requirements and electric charge and color conservation is given by
\cite{buch}
\begin{eqnarray}
{\cal L}_{\text{eff}}~  & =&  {\cal L}_{F=2} ~+~ {\cal L}_{F=0} 
~+~ \mbox{h.c.}
\; , 
\label{e:int}
\\
{\cal L}_{F=2}~  &=& g_{\text{1L}}~ \bar{q}^c_L~ i \tau_2~ 
\ell_L ~S_{1L}+ 
g_{\text{1R}}~ \bar{u}^c_R~ e_R ~ S_{1R} 
+ \tilde{g}_{\text{1R}}~ \bar{d}^c_R ~ e_R ~ \tilde{S}_1
+ g_{3L}~ \bar{q}^c_L~ i \tau_2~\vec{\tau}~ \ell_L \cdot \vec{S}_3 
\nonumber \\
&& +~
g_{2L}~ (V^L_{2\mu})^T~ \bar{d}^c_R{\gamma}^{\mu}i{\tau}_2l_L~ +~
g_{2R}~ \bar{q}^c_L{\gamma}^{\mu}i{\tau}_2e_R~ V^R_{2\mu}
~+~ \tilde{g}_{2L}~ (\tilde{V}^L_{2\mu})^T~
\bar{u}^c_R{\gamma}^{\mu}i{\tau}_2l_L
\; ,
\label{lag:fer}\\
{\cal L}_{F=0}~  &=& h_{\text{2L}}~ R_{2L}^T~ \bar{u}_R~ i \tau_2 ~
 \ell_L 
+ h_{\text{2R}}~ \bar{q}_L  ~ e_R ~  R_{2R} 
+ \tilde{h}_{\text{2L}}~ \tilde{R}^T_2~ \bar{d}_R~ i \tau_2~ \ell_L
+ h_{1L}~ \bar{q}_L{\gamma}^{\mu}l_L~  U_{1\mu}^L~
\nonumber \\
&&+~ 
h_{1R}~ \bar{d}_R{\gamma}^{\mu}e_R~ U_{1\mu}^R~
+\tilde{h}_{1R}~ \bar{u}_R{\gamma}^{\mu}e_R~ \tilde{U}_{1\mu}^R
+~ h_{3L}~ \bar{q}_L\vec{\tau}{\gamma}^{\mu}l_L~ \vec{U}_{3\mu}^L
\; ,
\label{eff} 
\end{eqnarray}
where $F=3B+L$, $q_L$ ($\ell_L$) stands for the left-handed quark (lepton)
doublet, and $u_R$, $d_R$, and $e_R$ are the singlet components of the
fermions. We denoted the charge conjugated fermion fields by $\psi^c =C
\bar\psi^T$ and we omitted in Eqs.\ (\ref{lag:fer}) and (\ref{eff}) the flavor
indices of the leptoquark couplings to fermions.  The leptoquarks $S_{1R(L)}$,
$\tilde{S}_1$, $U^{L(R)}_{1\mu}$, and $\tilde{U}^R_{1\mu}$ are singlets under
$SU(2)_L$, while $R_{2R(L)}$, $\tilde{R}_2$, $V^{R(L)}_{2\mu}$, and
$\tilde{V}^L_{2\mu}$ are doublets, and $S_3$ and $\vec{U}_{3\mu}^L$ are
triplets. The quantum numbers for all leptoquarks can be found, for instance,
in the last reference of \cite{fut:ee}. In this work, we denoted the Yukawa
couplings $h$ and $g$ by $\kappa$.


We can see from the above interactions that the main decay modes of
leptoquarks are into pairs $e^\pm q$ and/or $\nu_e q^\prime$, thus, their
signal is either a $e^\pm$ plus a jet, or a jet plus missing energy. However,
this is true provided the leptoquark masses are such that they can not decay
into another leptoquark belonging to the same multiplet and a vector boson.
Here we assumed that the leptoquarks belonging to the same multiplet are
degenerate in mass.  Furthermore, we implicitly assumed that the leptoquarks
couple only to the known particles, {\it i.e.} we do not consider the scenario
where the leptoquarks also couple to other new particles like charginos or
neutralinos in $R$--parity violating SUSY models.

In this work we considered only the $e^\pm q$ decay
mode and took into account the corresponding branching ratio.  We exhibit in
Table \ref{t:cor} the leptoquarks that can be analyzed using the final state
$e^\pm$ plus a jet, as well as, their decay products. As we can see from Eqs.\ 
(\ref{lag:fer}) and (\ref{eff}), only the leptoquarks $R^2_{2L}$,
$\tilde{R}^1_2$, $S^+_{3}$, $V^{2\mu}_2$, and $U_{3}^{+\mu}$ decay exclusively
into a jet and a neutrino.


There have been many searches for leptoquarks which, so far, led to negative
results.  Analyzing the decay of the $Z$ into a pair of on-shell leptoquarks,
the LEP experiments established a lower bound $ M_{\text{lq}} \gtrsim 44$ GeV
for scalar leptoquarks \cite{LEP,DELPHI}. Recently the LEP Collaborations
\cite{lepnew} used their $\sqrt{s}= 161$ and 172 GeV data to obtain the
constraint $M_{\text{lq}} \gtrsim 131$ GeV for leptoquarks coupling to first
family quarks and electrons.  The search for scalar (vector) leptoquarks
decaying exclusively into electron-jet pairs at the Tevatron constrained their
masses to be $M_{\text{lq}} \gtrsim 225$ (240) GeV \cite{PP}. Furthermore, the
experiments at HERA \cite{HERA} placed limits on their masses and couplings,
establishing that $M_{\text{lq}} \gtrsim 216-275$ GeV depending on the
leptoquark type and couplings.


Low-energy experiments also lead to strong indirect bounds on the couplings
and masses of leptoquarks, which can be used to define the goals of new
machines to search for these particles.  The main sources of indirect
constraints are:

$\bullet$ Leptoquarks give rise to Flavor Changing Neutral Current (FCNC)
processes if they couple to more than one family of quarks or leptons
\cite{shanker,fcnc}. In order to avoid strong bounds from FCNC, we assumed
that the leptoquarks couple to a single generation of quarks and a single one
of leptons. However, due to mixing effects on the quark sector, there is still
some amount of FCNC left \cite{leurer} and, therefore, leptoquarks that couple
to the first two generations of quarks must comply with some low-energy bounds
\cite{leurer}.

$\bullet$ The analyses of the decays of pseudoscalar mesons, like the pions,
put stringent bounds on leptoquarks unless their coupling is chiral -- that
is, it is either left-handed or right-handed \cite{shanker}.

$\bullet$ Leptoquarks that couple to the first family of quarks and leptons
are strongly constrained by atomic parity violation \cite{apv}.  In this case,
there is no choice of couplings that avoids the strong limits.

$\bullet$ The analyses of the effects of leptoquarks on the $Z$ physics
through radiative corrections lead to limits on the masses and couplings of
leptoquarks that couple to top quarks \cite{gb,jkm}.

As a rule of a thumb, the low-energy data constrain the masses of leptoquarks
to be larger than $0.5$--$1$ TeV when their Yukawa coupling is equal to the
electromagnetic coupling $e$ \cite{leurer,jkm,davi}. Therefore, our results
indicate that the LHC can not only confirm these indirect limits but also
expand them considerably.

\section{Signals and Backgrounds for the Production of Leptoquarks}
\label{sign}

In this work we focus our attention on the $s$--channel leptoquark production
via
\begin{equation}
pp ~\rightarrow q q ~\rightarrow~ \gamma q ~\rightarrow~ e^\pm q 
~\rightarrow \Phi_{\text{lq}}
~\rightarrow~ e^\pm q \; ,
\label{eq:nois}
\end{equation}
which leads to $e^\pm$--jet pairs with balanced transverse momenta, up to the
detector resolution.  In our evaluation of the subprocess cross section $e q
\to e q$, we included the irreducible SM background due to the $\gamma$ and
$Z$ exchange, treating properly its interference with the leptoquark diagrams;
see Fig.\ \ref{f:feyn}. The expressions for the subprocess cross sections
($\hat{\sigma}_{eq\to eq}$) are presented in Appendix A for all leptoquark
models.  The leptoquark production cross section is then obtained by folding
$\hat{\sigma}_{eq \to eq}$ with the quark ($f_{q/p}$) and $e^\pm$ ($f_{e/p}$)
distributions in the proton:
\begin{equation}
\sigma(pp \to e q X) = \int d x_e dx_q~ f_{e/p}(x_e)~
 f_{q/p}(x_q)~ \hat{\sigma}_{eq\to eq}(\hat{s})\; ,
\end{equation}
where the subprocess center--of--mass energy ($\sqrt{\hat{s}}$) is related to
the $pp$ one ($\sqrt{s}$) by $\hat{s} = x_e x_q s$.

The distribution of $e^\pm$ in the proton is given by
\begin{equation}
f_{e/p}(x_e) = \int^1_{x_e} \frac{dz}{z} f_{e/\gamma} (z) 
  f_{\gamma/p}\left(  \frac{x_e}{z} \right)  \; ,
\label{fpe}
\end{equation}
with $f_{\gamma/p}$ being the distribution of photons in the proton and the
splitting rate of $\gamma$ into $e^+e^-$ pairs given by \cite{kessler}
\begin{equation}
f_{e/\gamma}(z) = \frac{\alpha}{2\pi}~ \left [ z^2 + (1-z)^2 \right ]~
 \log\left( \frac{Q^2}{m_e^2} \right ) \; .
\end{equation}

We chose the scale $Q^2 = M^2_{lq}$ and denoted by $m_e$ ($M_{\text{lq}}$) the
electron (leptoquark) mass.  There are two possibilities for radiation photons
of the proton: either the photons are radiated by the proton as a whole and it
does not break off, or quarks radiate the photons and the proton fragments.
Since this last mechanism leads to a larger photon flux, we considered only it
in our analyses. In this case, the photon distribution in the proton is
\cite{zerwas}
\begin{equation}
f_{\gamma/p}(x_\gamma) = \frac{\alpha}{2\pi}~ \log{\left(\frac{Q^2}
{m^2_{q}}\right)} \frac{1}{x_{\gamma}} \int^1_{x_\gamma}
\frac{dz}{z} \left[ 1 + (1-x_\gamma/z)^2 \right] F_2(z,Q^2)\; ,
\label{fqg2} 
\end{equation}
with $F_2(z,Q^2)$ being the structure function of a quark inside the proton
summed over all the quark flavors including the electric charge factors.

There are many SM processes that lead to the production of $e^\pm$ and jets.
Since some of them give rise to more than one jet or $e^\pm$, we can enhance
the signal demanding the presence of a single $e^\pm$ and a single jet in the
central region of the detector.  This requirement eliminates dangerous
backgrounds like the pair production of electroweak gauge bosons or top quarks
\cite{thais}. Nevertheless, there are further backgrounds for the leptoquark
search like the scattering of $e^\pm$ and (anti) quarks with flavors different
of the leptoquark ones; see Fig.\ \ref{f:feyn}.

In actual experiments, the observed signal events will not possess a balance
between the $e^\pm$ and jet transverse momenta due to the experimental
detector resolution.  Consequently, we must analyze backgrounds like the
$W$--jet production, where the $W$ decays into a pair electron--neutrino.
Moreover, we should also consider the SM production of $Z$--jet pairs with the
$Z$ decaying into a $e^+e^-$ pair and one of the $e^\pm$ escaping undetected.

We mimicked the experimental resolution of the electromagnetic and hadronic
calorimeters by smearing the final state quark and lepton energies according
to
\begin{eqnarray}
\left. \frac{\delta E}{E}\right|_{em} &=& \frac{0.02}{\sqrt{E}} \oplus 0.005
\hskip 0.5cm \hbox{  electromagnetic} \; ,
\\
\left.\frac{\delta E}{E}\right|_{had} &=& \frac{0.6}{\sqrt{E}} \oplus 0.03
\hskip 0.5cm \hbox{  hadronic} \; .
\end{eqnarray}
Angles were smeared in a cone with
\begin{eqnarray}
\left. \delta\theta\right|_{em} &=& 10\;\; {\rm mrad} \hskip 0.5cm
\hbox{  electromagnetic} \; ,
\\
\left. \delta\theta\right|_{had} &=& 15\;\; {\rm mrad} \hskip 0.5cm
\hbox{  hadronic} \; .
\end{eqnarray}

\begin{figure}
\centering\leavevmode
\psfig{file=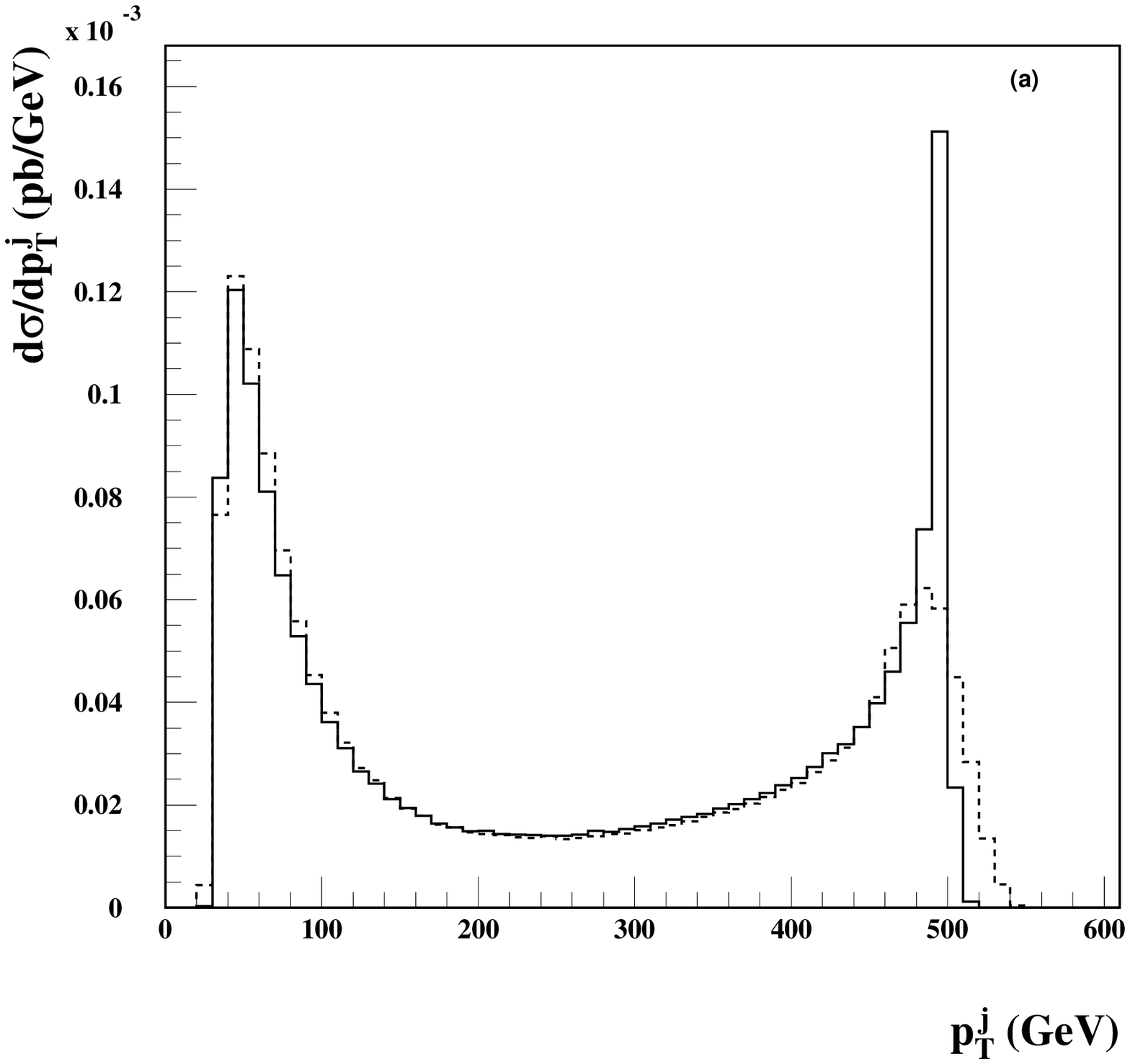,height=8cm,width=8cm}
\psfig{file=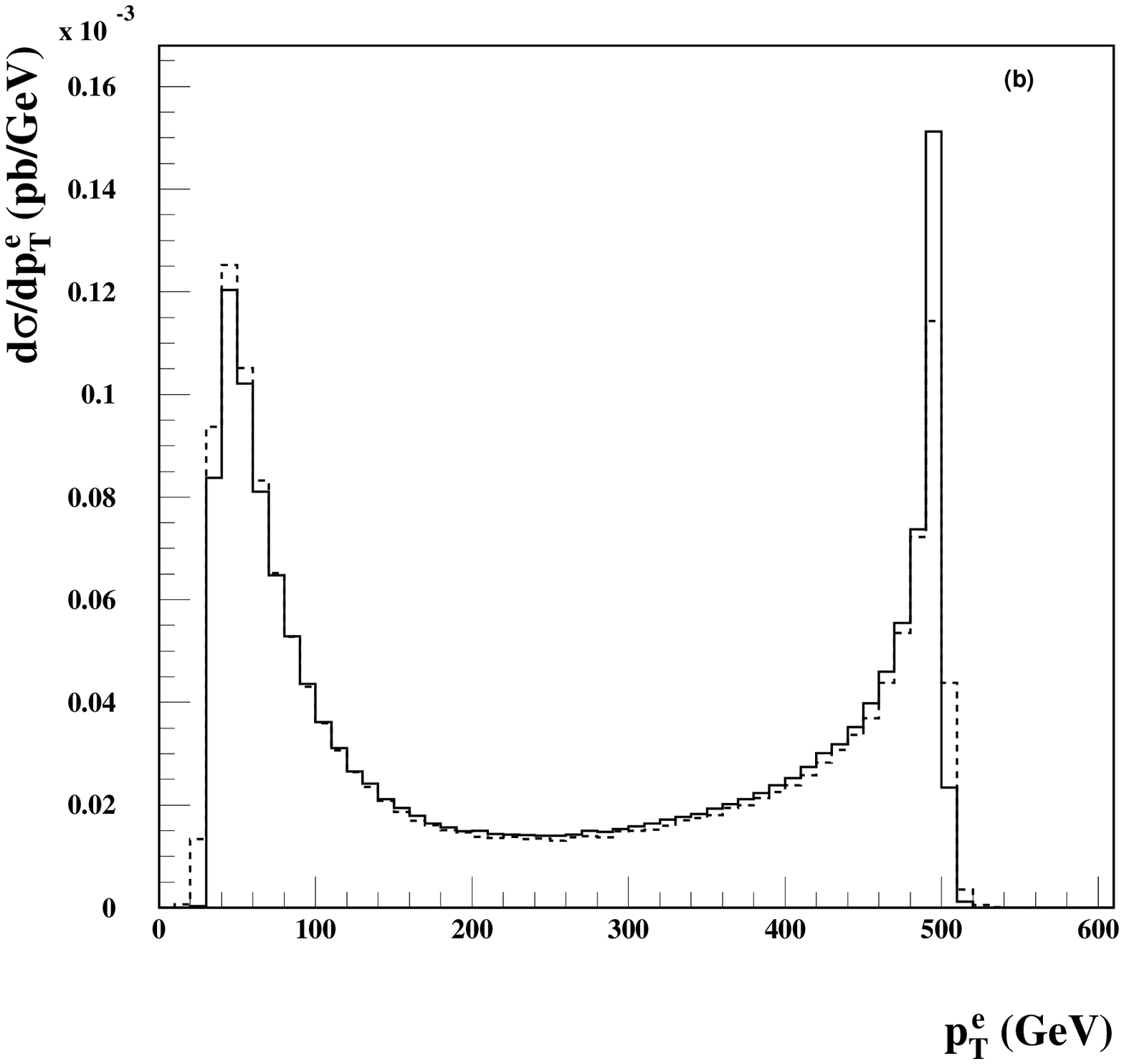,height=8cm,width=8cm}
\caption{$p_T$ distributions of the jet (a) and $e^\pm$ (b) originating from 
  leptoquarks before (solid line) and after (dashed line) applying the
  calorimeter resolution for $M_{\text{lq}} = 1$ TeV and $\kappa = 0.3$. We
  imposed the pseudorapidity cuts $|y_{e^\pm ,j}| < 3.5$ and required the
  $e$--jet invariant mass to be in the range $|M_{\text{lq}}\pm 40|$ (GeV).}
\label{pt:ejet}
\end{figure}

We show in Figs.\ \ref{pt:ejet}a and \ref{pt:ejet}b the typical behavior of
the $p_T$ distributions of the jet and $e^\pm$ before (solid line) and after
(dashed line) applying the calorimeter resolution for the process
(\ref{eq:nois}), including all $eq \to eq$ irreducible backgrounds; see Fig.\ 
\ref{f:feyn}. In these figures, we assumed a $S_{1L}$ leptoquark with
$M_{\text{lq}} = 1$ TeV and $\kappa = 0.3$, and we also required that
$|y_{e^\pm ,j}| < 3.5$ and the invariant mass of the $e$--$j$ pair ($M_{ej}$)
to be in the range $|M_{\text{lq}}\pm 40|$ GeV.  The peak around $p_T =
M_{\text{lq}}/2$ is due to the leptoquark production while the low $p_T$ peak
is associated to the SM backgrounds. This feature of the $p_T$ spectrum
provides an efficient way to separate the leptoquark signal from backgrounds.
Moreover, the calorimeter resolution broadens the peak associated to the
signal and increases slightly the low--$p_T$ peak associated to the
$t$--channel backgrounds without changing significantly the total cross
section.

\begin{figure}
\centerline{
\psfig{file=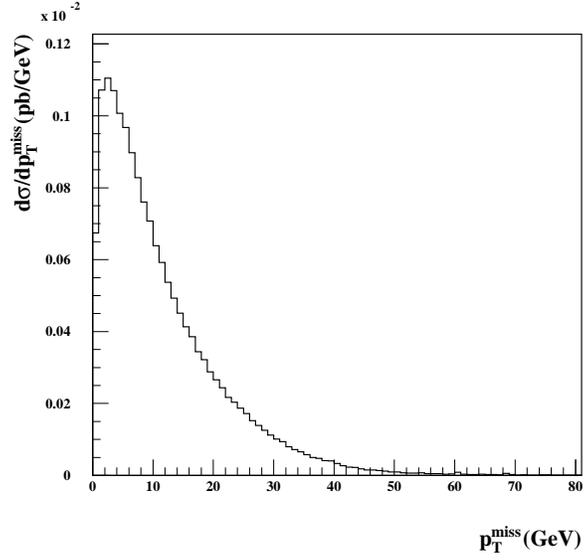,height=8cm,width=8cm}}
\caption{Missing $p_T$ distribution in the process (\protect\ref{eq:nois})
  due to the calorimeter resolution.  We assumed the same parameters and cuts
  used in Fig.\ \protect\ref{pt:ejet}.}
\label{pt:miss}
\end{figure}

We show in Fig.\ \ref{pt:miss} the missing $p_T$ spectrum originated from the
smearing of the momenta of the final state jet and $e^\pm$ in process
(\ref{eq:nois}), using the same parameters and cuts of Fig.\ \ref{pt:ejet}. As
expected, the missing $p_T$ distribution is peaked at small values, being
negligible for missing $p_T$'s larger than 70 GeV.  Moreover, the missing
$p_T$ in the signal events should be parallel to the total $p_T$ of the
$e^\pm$--jet system since the main effect of the experimental resolution is to
alter the magnitude of the measured transverse momenta.

\begin{figure}
\centering\leavevmode
\psfig{file=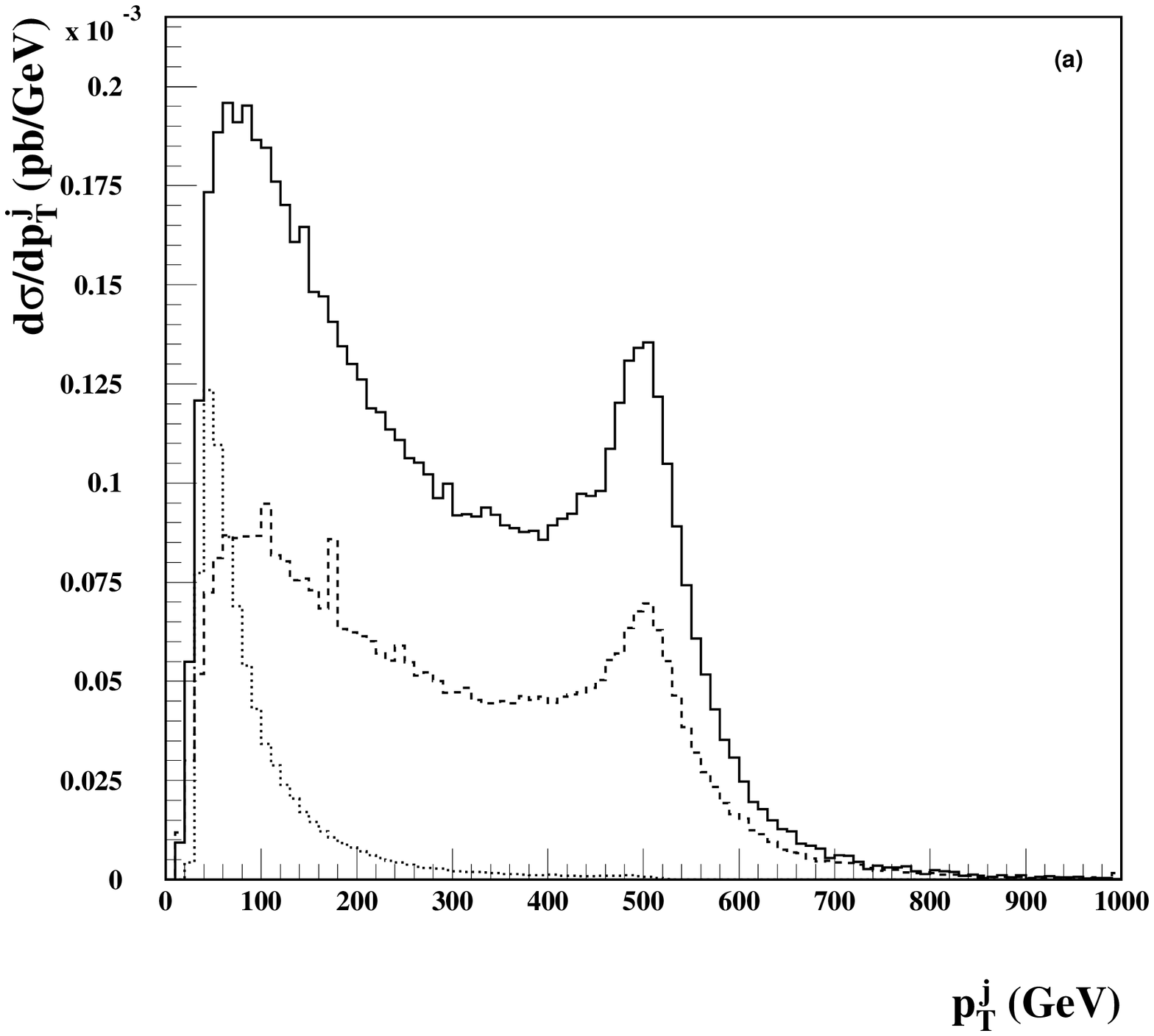,height=8cm,width=8cm}
\psfig{file=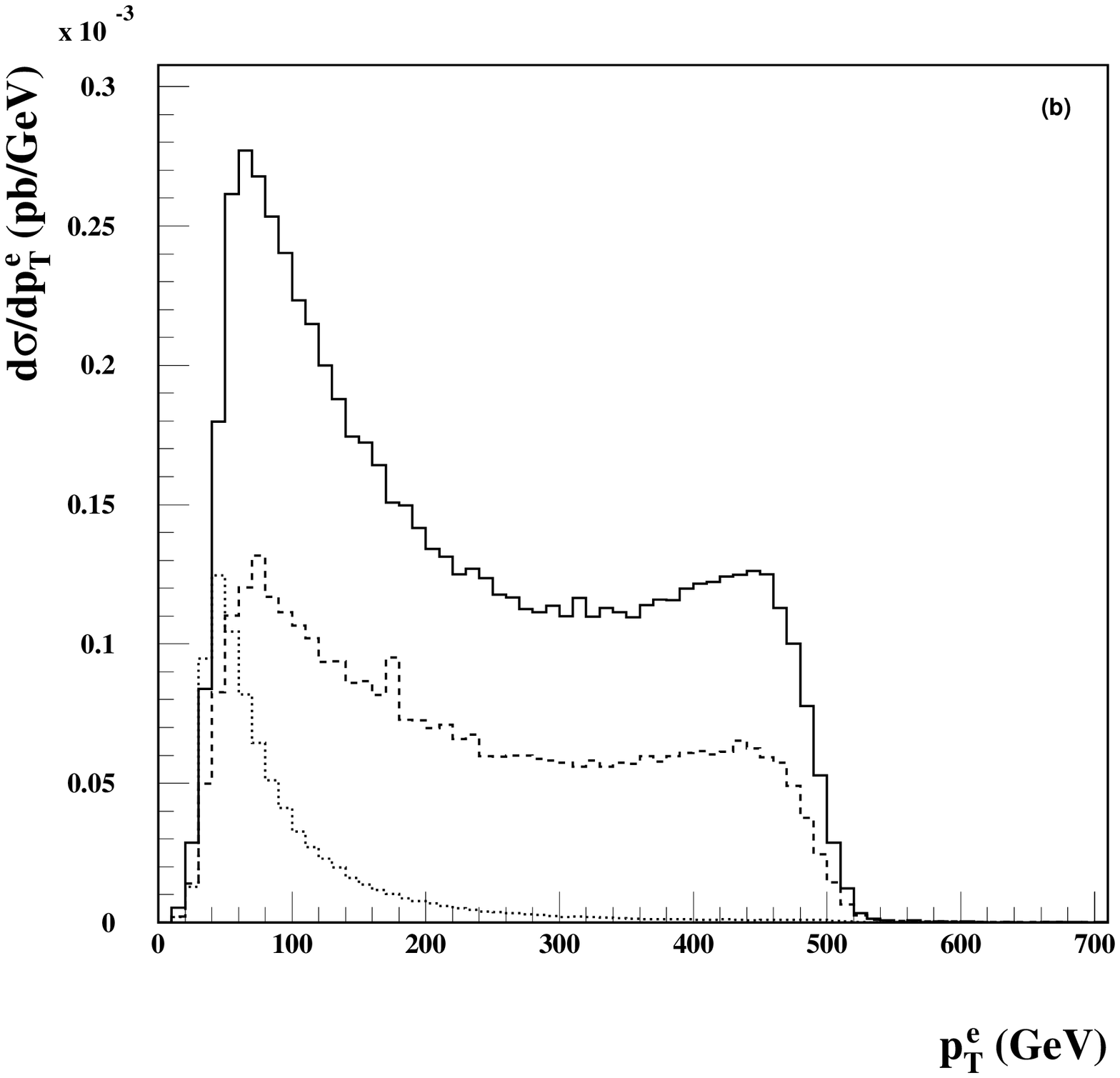,height=8cm,width=8cm}
\psfig{file=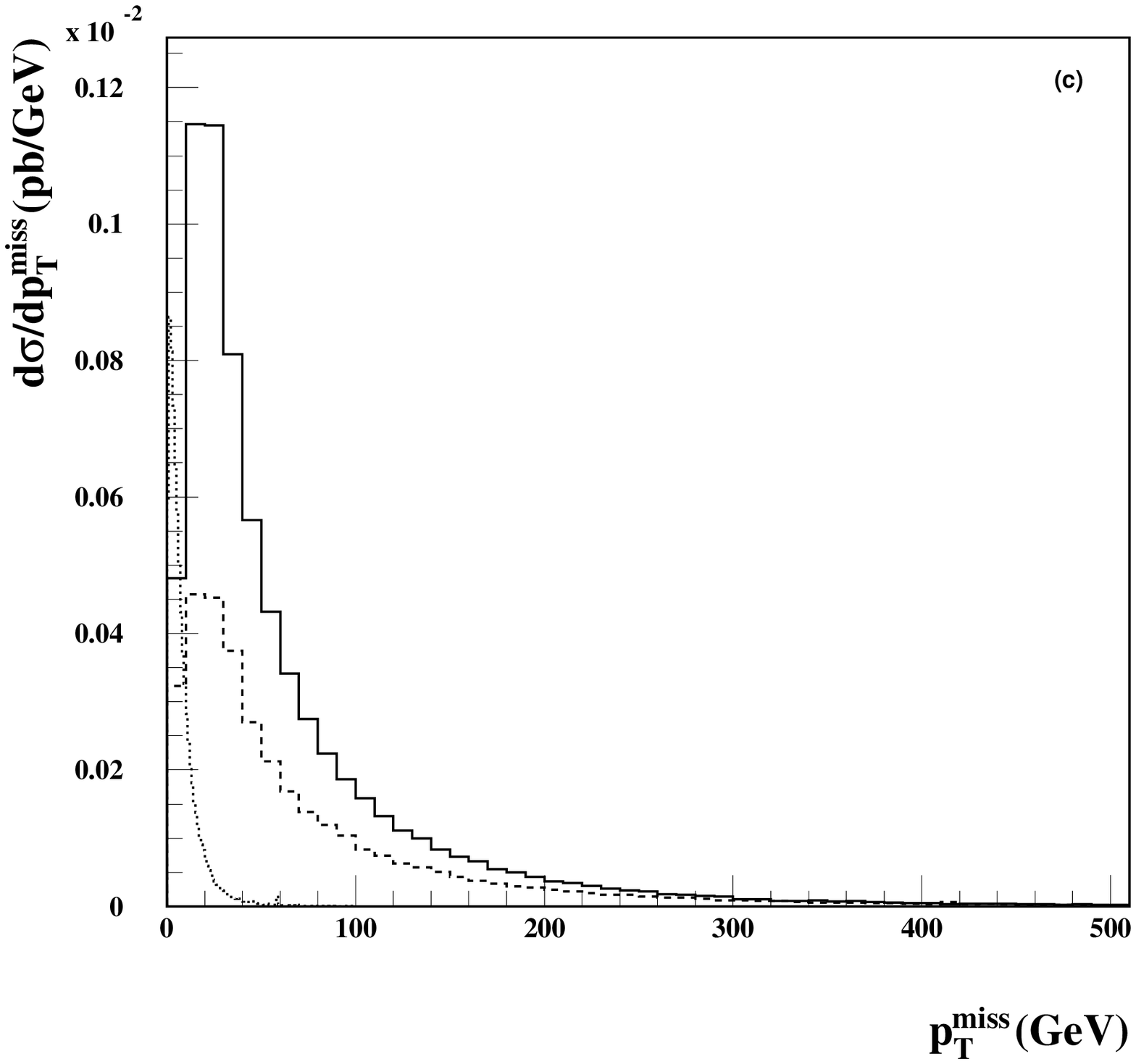,height=8cm,width=8cm}
\caption{Jet (a) , $e^\pm$ (b),  and  missing (c) $p_T$ spectrum 
  originating from the $W$--jet (solid line), $Z$--jet (dashed line), and
  $eq\to eq$ with $\kappa=0$ (dotted line) backgrounds after applying the
  calorimeter resolution. We imposed the same cuts used in Fig.\ 
  \protect\ref{pt:ejet} and required $p_T^{j,e} > 10$ GeV.}
\label{pt:bg}
\end{figure}

We present in Fig.\ \ref{pt:bg} several distributions for the backgrounds
described above which should be contrasted with the signal ones. Fig.\ 
\ref{pt:bg}a (b) contains the $p_T$ spectrum of the jet ($e^\pm$) coming from
the backgrounds after we applied the same cuts used in Fig.\ \ref{pt:ejet} 
and required $p_T^{j,e} > 10$ GeV to avoid divergences due to gluons. As
we can see, the backgrounds are peaked at low transverse momentum of the jet
($e^\pm$) with the largest background being $W$--jet production. Despite the
lack of a $s$--channel resonance, the $p_T$ spectra peak around
$M_{\text{lq}}/2$ due to the $e^\pm$--jet invariant mass cut.  The missing
$p_T$ distribution of the background, shown in Fig.\ \ref{pt:bg}c, reaches its
maximum around 40 GeV and extends up to 300 GeV approximately, possessing a
larger fraction of events at large missing $p_T$ than the signal.

Taking into account the above features of the signal and backgrounds we
imposed the following set of cuts in order to enhance the signal and suppress
the backgrounds:

\begin{itemize}
  
\item [(C1)] The first requirement is that the jet and $e^\pm$ are in the
  pseudo-rapidity interval $|y| < 3.5$.
        
\item [(C2)] We also demanded the events to have the $e$-jet invariant mass in
  the range $| M_{\text{lq}} \pm \Delta M|$ with $\Delta M$ given in
  Table~\ref{bins}.
  
\item [(C3)] We veto events exhibiting an extra $e^\pm$ (or parton) in the
  region $|y| < 3.5$. This cuts reduces backgrounds like $t\bar{t}$ production
  which exhibit many more $e^\pm$ or jets in the central rapidity region.
  
\item [(C4)] The $e^\pm$ and jet should have $p_T > p_T^{min}$ with
  $p_T^{min}$ given in Table~\ref{bins}.
      
\item [(C5)] We apply a cut on the missing $p_T$ requiring its value be lower
  than those in Table~\ref{bins}.
  
\item [(C6)] Finally, we require that the cosine of the angles between the
  direction of missing $p_T$ and the $p_T$ of $e^\pm$ and jet to be larger
  than 0.94.

\end{itemize}

In principle we should also require the $e^\pm$ to be isolated from hadronic
activity in order to reduce the QCD backgrounds. Nevertheless, it was shown in
Ref.\ \cite{thais} that this cut does not further suppress the background
after we apply the cut C4.


\section{Results}
\label{resu}

We present in Table~\ref{uncut:signal}, as an illustration, the total cross
section for producing pairs $e^-$--jet and $e^+$--jet, applying two different
sets of cuts to the smeared final state momenta. We assumed in this table that
$M_{\text{lq}}= 1$ TeV and $\kappa = 0.3$. We denoted the irreducible $eq \to
eq$ background by $\sigma_{bg}$, which was obtained setting $\kappa=0$. The
cuts C3--C6 reduce the backgrounds by two or more orders of magnitude while
the efficiency for the signal is of the order of $10$--$20$\% depending on the
leptoquark type.  As we can see, the most important background is the $W$--jet
production, which is larger then the $Z$--jet and irreducible backgrounds by a
factor of roughly 20. Moreover, we verified using ISAJET \cite{isajet}
 that the $t \bar{t}$
production background is effectively reduced by our cuts, specially C3, being
it negligible face to the $W$--jet production. Therefore, the LHC reach in
$\kappa$ and $M_{\text{lq}}$ will be controlled by the $W$--jet background.

Leptoquarks of the type $F=2$ couple to pairs $e^- q$, and consequently are
more copiously produced in $s$--channel processes leading to the final state
$e^-$--jet than in reactions leading to $e^+$--jet since there are more quarks
than anti-quarks in the proton. For $F=0$ leptoquarks the situation is the
opposite since they couple to $e^- \bar{q}$ pairs. As expected, the results
shown in Table~ \ref{uncut:signal} agree with these arguments.  Furthermore,
using these features of $F=0$ and $F=2$ leptoquarks, we can differentiate
between them simply by counting the number of leptoquark events with electrons
and positrons in the final state.

\begin{figure}
\centering\leavevmode
\psfig{file=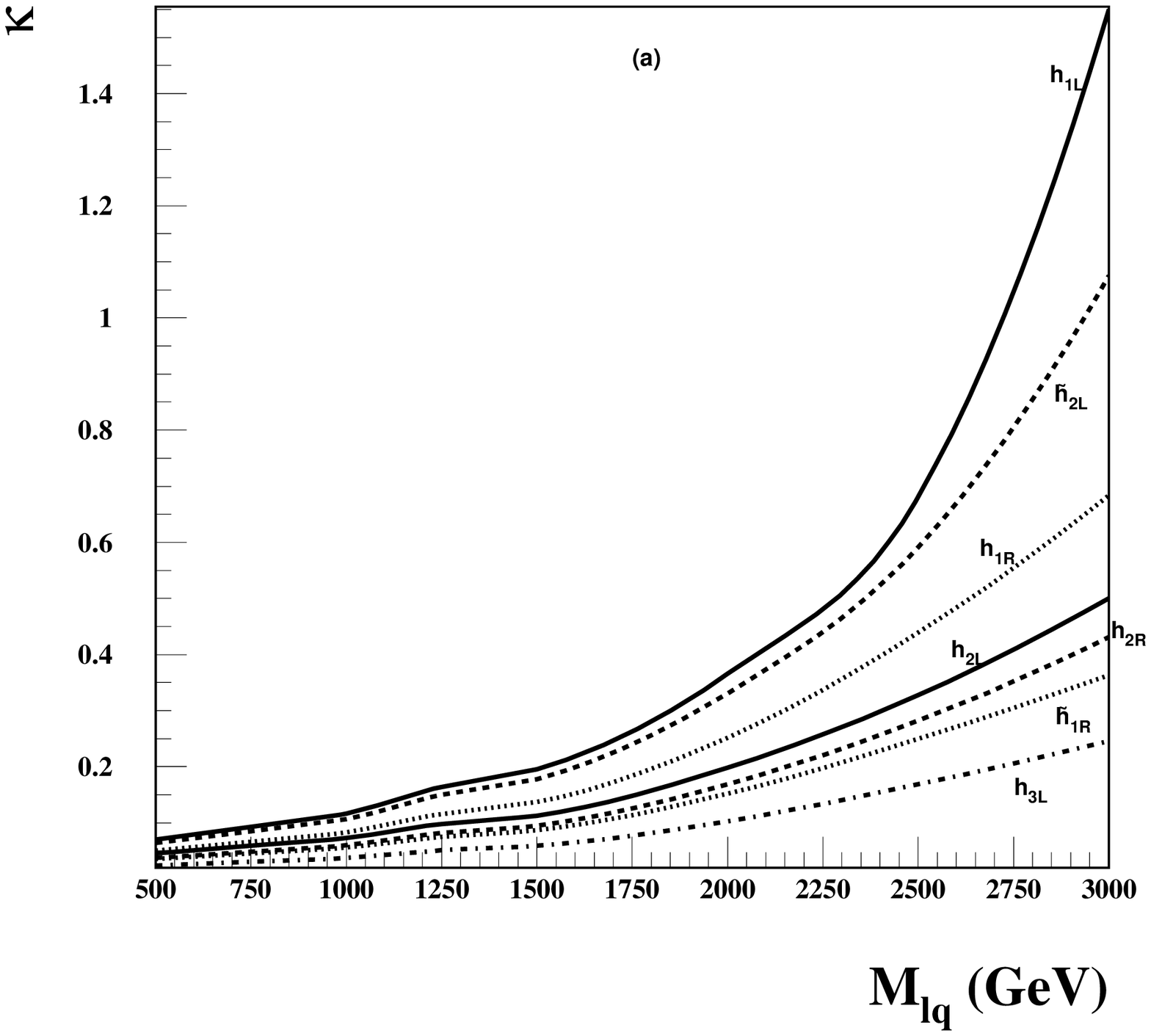,height=8cm,width=8cm} 
\psfig{file=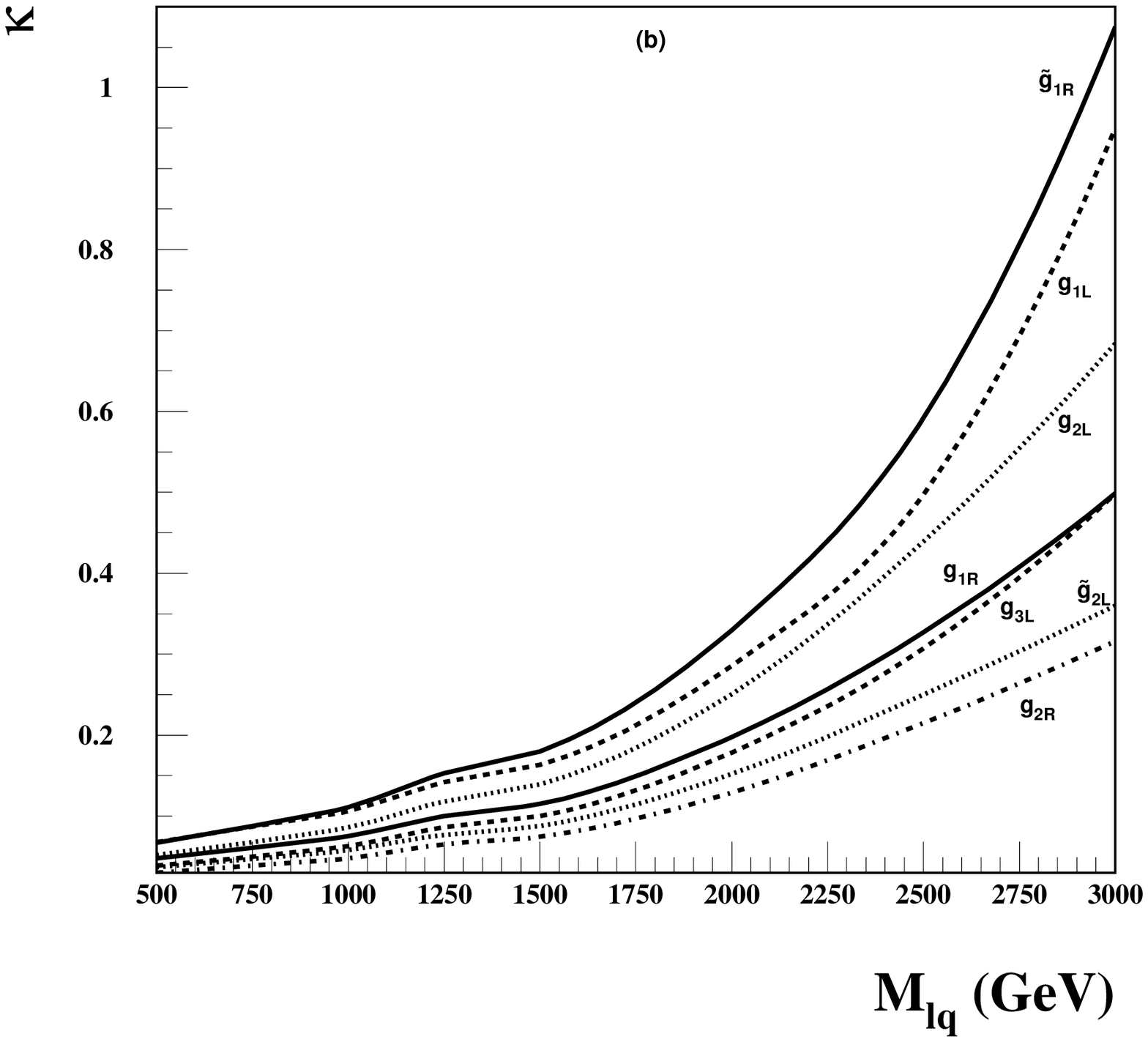,height=8cm,width=8cm}
\caption{99.73\% CL excluded regions in the plane $\kappa$--$M_{\text{lq}}$ 
  from negative searches of single production of leptoquarks with $F=0$ (a)
  and $F=2$ (b) for an integrated luminosity of 100 fb$^{-1}$.}
\label{lim}
\end{figure}

In order to obtain the LHC attainable limits on leptoquarks we employed the
final state $e^-$--jet ($e^+$--jet) for $F=2$ ($F=0$) leptoquarks since this
topology possesses the largest signal cross section. Figs.\ \ref{lim}a and
\ref{lim}b contain the regions in the plane $\kappa \times M_{\text{lq}}$ that
can be excluded at the 99.73\% CL ($3\sigma$ level) from negative single
leptoquark searches at the LHC for an integrated luminosity of 100 fb$^{-1}$.
As we can see, the LHC will be able to discover leptoquarks with masses of at
least $2$ TeV for leptoquark Yukawa couplings of the electromagnetic strength
($\kappa = 0.3$).  The $U_{3\mu}$ leptoquarks will exhibit the tightest bounds
while the $U_{1\mu}$ leptoquarks will possess the loosest limits.  Moreover,
our results are comparable with those presented in \cite{thais} for leptoquark
searches using the processes (\ref{eq:sin})--(\ref{eq:gg}).

\begin{figure}
\centerline{
\psfig{file=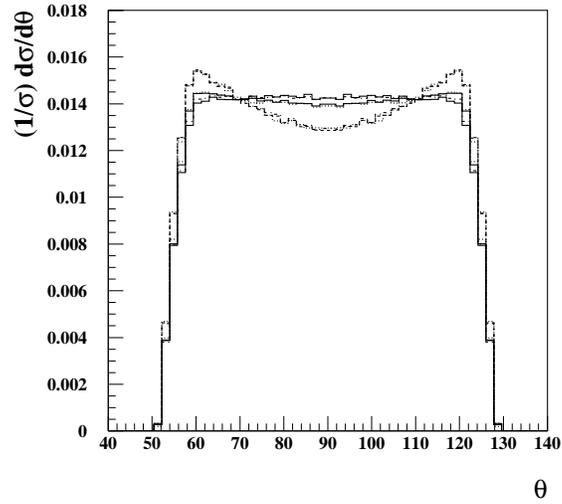,height=8cm,width=8cm}}
\caption{Normalized polar angular distributions of the electron in the
  leptoquark rest frame, including the $W$ + jet background. We assumed that
  $\kappa=0.3$ and $M_{\text{lq}}=1$ TeV and imposed the cuts C1--C6. The
  flatter lines correspond to the scalar $F=2$ leptoquarks while the peaked
  ones to the vector $F=2$ ones.}
\label{angular}
\end{figure}

We should also study the capability of the LHC to unravel the properties of
leptoquarks in the event a signal is observed. As discussed above, the ratio
of signal events in the channels $e^+$--jet and $e^-$--jet can discriminate
between $F=0$ and $F=2$ leptoquarks.  In order to learn more about the
leptoquark giving rise to the signal, we should also study kinematical
distributions. For instance, the $e^\pm$ polar angle distribution for scalars
and vectors are distinct in the leptoquark rest frame. As an example, we show,
in Fig.\ \ref{angular}, the $e^-$ normalized polar angle spectrum, including
the backgrounds, for all $F=2$ leptoquarks. In this figure, we assumed
$\kappa=0.3$ and $M_{\text{lq}}=1$ TeV and applied the cuts C1--C6.  As
expected, the distributions of scalar and vector leptoquarks are different,
being the scalar distribution flatter. Nevertheless, the discover of the
leptoquark spin will only be possible provided there will be enough events to
render a statistical meaning to the angular distribution.

\begin{figure}
\centering\leavevmode
\psfig{file=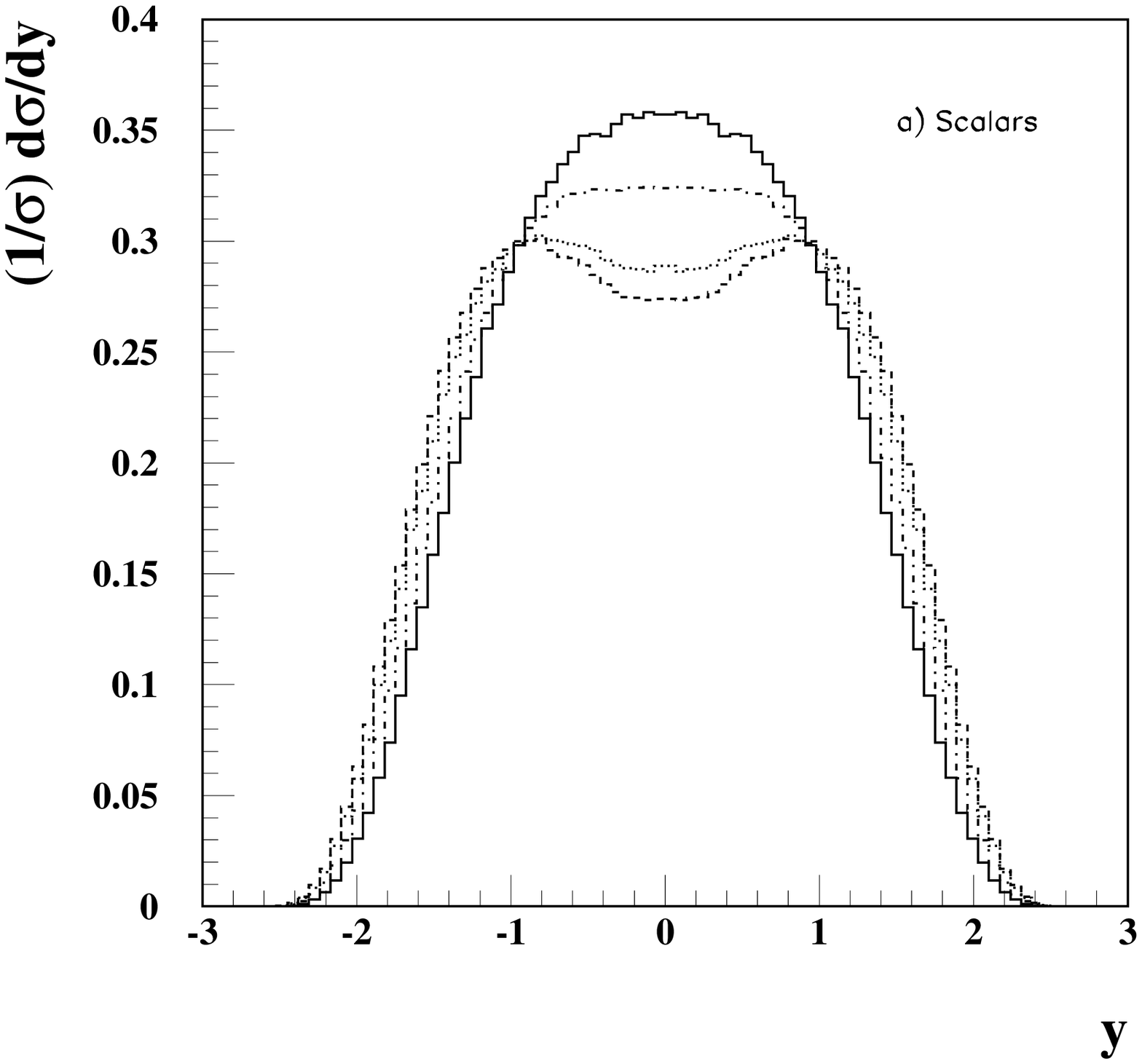,height=8cm,width=8cm}
\psfig{file=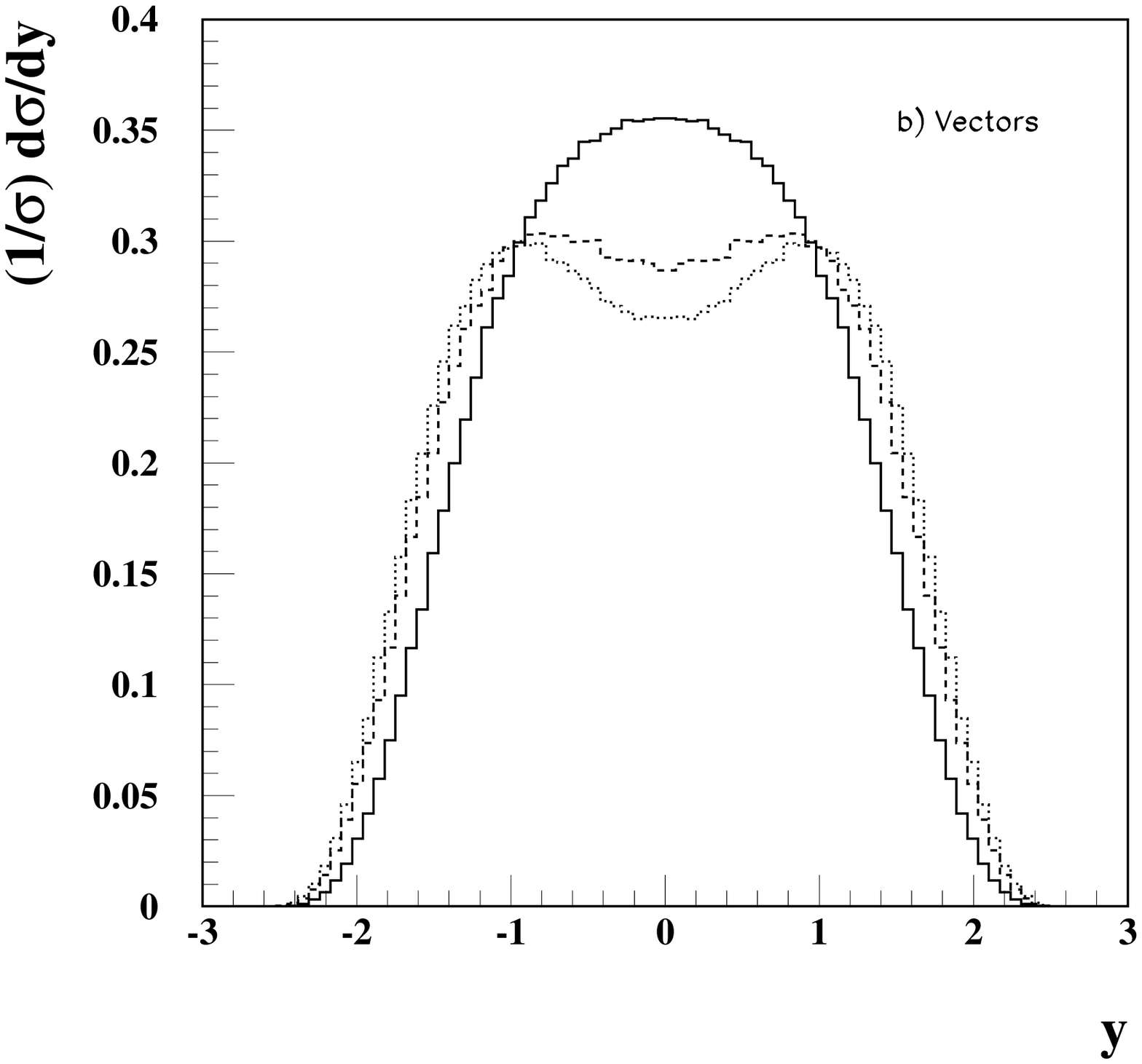,height=8cm,width=8cm}
\caption{Normalized pseudorapidity distributions of the leptoquarks,
  including the $W$ + jet background. We show the results for all types of
  leptoquarks with $F=2$, considering $\kappa=0.3$ and $M_{\text{lq}}=1$ TeV.
  In (a) the solid line is for $\tilde{S}_1$, the dashed line stands for
  $S_{1R}$, the dotted line represents $S_{1L}$, and the dashed-dotted line is
  for $S_{3BC}$. In (b) the solid line is for $V^\mu_{2L}$ the dashed line
  represents $V^\mu_{2R}$, and the dotted line stands for
  $\tilde{V}^\mu_{2}$.}
\label{rap}
\end{figure}

We can distinguish leptoquarks that couple to $u$ or $d$ quarks analyzing the
leptoquark pseudo-rapidity distribution in the lab frame because leptoquarks
coupling to $u$ quarks are produced at larger rapidities than the ones
coupling to $d$'s.  In Fig.\ \ref{rap}, we show the normalized distributions
after cuts for the pseudo-rapidity of scalar (a) and vector leptoquarks (b)
with $F=2$, where the backgrounds were added to the signal.  We can see three
distinct curves in Fig.\ \ref{rap}a: the largest distribution at central
pseudo-rapidities is due to $\tilde{S}_1$, which couples only to $d e^-$; the
leptoquark triplet $S_3$ couples to $de^-$ and $ue^-$ and gives rise to the
curve in the middle; the curves peaked at higher rapidities originate from the
production of $S_{1R}$ and $S_{1L}$ which couple exclusively to $e^- u$. Since
$S_{1R}$ possesses a larger branching ratio into $e^- u$ than $S_{1L}$, its
distribution is larger at high rapidities and less affected by the
backgrounds.  The situation is analogous for vector leptoquarks, as can be
seen from Fig.\ \ref{rap}b.


\section{Conclusions}
\label{conc}

The discover of leptoquarks is without any doubt a striking signal for the
existence of life beyond the standard model. In this work we demonstrated that
the search for leptoquarks in the process $ pp \to e^\pm q \to$ leptoquark
$\to e^\pm q$ at the LHC will be able to exclude leptoquarks with masses
smaller than 2--3 TeV for Yukawa couplings of the order of the
electromagnetic ones and an integrated luminosity of 100 fb$^{-1}$. Our
results are summarized in Table~\ref{tab:lim}. It is important to notice that
our bounds are comparable to the ones coming from the reactions
(\ref{eq:sin})--(\ref{eq:gg}) \cite{thais}. Therefore, it will be possible to
make a cross check between the different channels and to improve the bounds
combining them. Furthermore, the LHC will be able not only to confirm the
present indirect limits on leptoquarks but also to expand them considerably.

If a leptoquark signal is observed at the LHC, we showed that we can determine
whether the leptoquark is $F=0$ or $F=2$ by counting the number of events in
the $e^-$--jet and $e^+$--jet final states.  Moreover, the spin of the
leptoquark can also be established from the $e^\pm$ polar angle distribution
in the leptoquark rest frame provided there are enough events for this
distribution to be meaningful.  We can even determined which leptoquark
multiplet was produced by studying the rapidity spectrum of the leptoquarks in
the lab frame. Finally, knowing the type of leptoquark we can estimate
$\kappa$ using the size of the cross section, and consequently determine all
leptoquark parameters. Once again, the pair and single production of
leptoquarks via (\ref{eq:sin})--(\ref{eq:gg}) can be used to confirm the
leptoquark properties.


\acknowledgments

One of us (O.J.P.E.) would like to thank the kind hospitality of the Institute
for Elementary Particle Research, University of Wisconsin--Madison, where the
part of this work was done. P.G.M. would like to thank Xerxes Tata for useful
discussion. This work was partially supported by Conselho Nacional de
Desenvolvimento Cient\'{\i}fico e Tecnol\'ogico (CNPq), Funda\c{c}\~ao de
Amparo \`a Pesquisa do Estado de S\~ao Paulo (FAPESP), by the University of
Wisconsin Research Committee with funds granted by the Wisconsin Alumni
Research Foundation, and by the U.S.\ Department of Energy under Grant
No.~DE-FG02-95ER40896 and DE-FG03-94ER40833.


\appendix
\section{Subprocess Cross Sections}

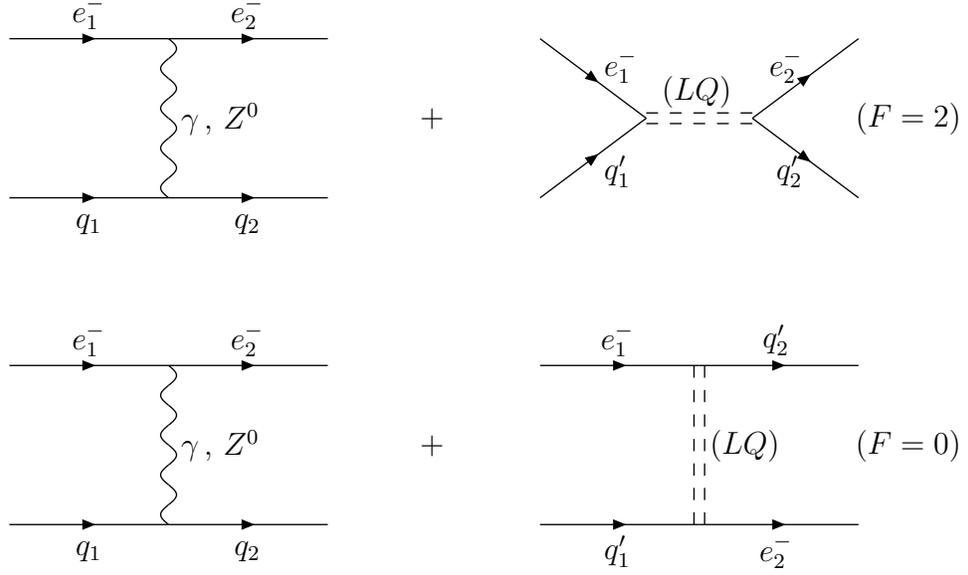
\begin{figure}
\vskip 36pt
\centerline{
\begin{picture}(320,80)(-15,-15)
\ArrowLine(0,70)(60,70)
\ArrowLine(0,10)(60,10)
\ArrowLine(60,70)(120,70)
\ArrowLine(60,10)(120,10)
\Photon(60,70)(60,10){3}{4}
\Text(30,80)[c]{$e^-_1$}
\Text(90,80)[c]{$e^-_2$}
\Text(30,0)[c]{$q_1$}
\Text(90,0)[c]{$q_2$}
\Text(65,40)[l]{$\gamma\,,\,Z^0$}
\Text(160,40)[c]{$+$}
\ArrowLine(200,70)(240,40)
\ArrowLine(200,10)(240,40)
\DashLine(240,42)(280,42){5}
\DashLine(240,38)(280,38){5}
\ArrowLine(280,40)(320,70)
\ArrowLine(280,40)(320,10)
\Text(225,60)[l]{$e^-_1$}
\Text(225,20)[l]{$q_1^{\prime}$}
\Text(260,50)[c]{$(LQ)$}
\Text(300,60)[r]{$e^-_2$}
\Text(300,20)[r]{$q_2^{\prime}$}
\Text(320,40)[l]{$(F=2)$}
\end{picture}
}
\vskip 1.5cm 
\centerline{
\begin{picture}(320,80)(-15,-15)
\ArrowLine(0,70)(60,70)
\ArrowLine(0,10)(60,10)
\ArrowLine(60,70)(120,70)
\ArrowLine(60,10)(120,10)
\Photon(60,70)(60,10){3}{4}
\Text(30,80)[c]{$e^-_1$}
\Text(90,80)[c]{$e^-_2$}
\Text(30,0)[c]{$q_1$}
\Text(90,0)[c]{$q_2$}
\Text(65,40)[l]{$\gamma\,,\,Z^0$}
\Text(160,40)[c]{$+$}
\ArrowLine(200,70)(260,70)
\ArrowLine(200,10)(260,10)
\ArrowLine(260,10)(320,10)
\ArrowLine(260,70)(320,70)
\DashLine(258,70)(258,10){5}
\DashLine(262,70)(262,10){5}
\Text(230,80)[c]{$e^-_1$}
\Text(290,80)[c]{$q^{\prime}_2$}
\Text(230,0)[c]{$q^{\prime}_1$}
\Text(290,0)[c]{$e^-_2$}
\Text(265,40)[l]{$(LQ)$}
\Text(320,40)[l]{$(F=0)$}
\end{picture}
}
\caption{Feynman diagrams that  contribute to the process $e^-\,q \to e^-\,q$, 
  with $q_i = u,d,s,c$ and $q_i^{\prime} = u,d$. We denoted the scalar and
  vector leptoquarks by $LQ$. }
\label{f:feyn}
\end{figure}

Here we collect the non--polarized differential cross sections in the
center-of-mass frame for the processes $e^{\pm}\,q~(\bar{q}) \rightarrow
e^{\pm}\,q~(\bar{q})$, including scalar and vector leptoquarks with $F=0$ and
$F=2$. The Feynman diagrams contributing to these reactions are displayed in
Fig.\ \ref{f:feyn}. The differential cross section for the process $e^-\,q
\rightarrow e^-\,q$ with the contribution of a $F=2$ scalar leptoquark is
given by
\begin{eqnarray}
\left(\frac{d\hat{\sigma}}{d\cos\theta}\right)_{e^-\,q \to e^-\,q}^{S}
&= \frac{1}{32\pi\hat{s}}
\,& \left[ |{\cal M}_{\gamma}|^2 + |{\cal
    M}_{Z^0}|^2 + |{\cal M}_S|^2 + 2{\rm Re}({\cal M}_{\gamma Z^0}) \right. 
\nonumber \\
&+& \left. 2{\rm Re}({\cal M}_{\gamma S}) + 2{\rm Re}({\cal M}_{Z^0
    S}) \right] \; ,
\label{dsdcosesc}
\end{eqnarray}
with
\begin{eqnarray} 
|{\cal M}_{\gamma}|^2 &=& 2Q_q^2e^4~\frac{\hat{s}^2 + 
\hat{u}^2}{\hat{t}^2} \; , 
\label{mg2} \\
|{\cal M}_{Z^0}|^2 &=& 2(G_F m_{Z^0}^2)^2\, \frac{1}{(\hat{t} - m_{Z^0}^2)^2}\,
\times
\nonumber \\
&& \left[(R_e^2R_q^2 + L_e^2L_q^2)\hat{s}  + (R_e^2 L_q^2 +
L_e^2R_q^2)\hat{u} \right] \; ,
\label{mz2} \\
|{\cal M}_{S}|^2 &=& \frac{\kappa^4}{4}\frac{\hat{s}^2}{(\hat{s}
- m_S^2)^2 + m_S^2\Gamma_S^2} \; , 
\label{ms2} \\
2{\rm Re}({\cal M}_{\gamma Z^0}) &=& -\frac{4G_F m_{Z^0}^2 Q_q e^2}{\sqrt{2}} 
\frac{1}{\hat{t}(\hat{t} - m_{Z^0}^2)} \, \times
\nonumber \\
&&\left[(R_eR_q + L_eL_q)\hat{s}^2 +
 (R_eL_q + L_eR_q) \hat{u} \right] \; , 
\label{mgz} \\
2{\rm Re}({\cal M}_{\gamma S}) &=& Q_q e^2\kappa^2\,
\frac{\hat{s}^2(\hat{s} - m_S^2)}
{\hat{t}\left[ (\hat{s} - m_S^2)^2 + m_S^2 \Gamma_S^2 \right]} \; ,
\label{mgs} \\
2{\rm Re}({\cal M}_{Z^0 S}) &=& - \frac{2 G_F m_{Z^0}^2
  \lambda^2_{L/R}}{\sqrt{2}} 
\frac{\hat{s}^2\,(\hat{s} - m_S^2)}{(\hat{t} - m_{Z^0}^2)\left[ 
(\hat{s} - m_S^2) + m_S^2 \Gamma_S^2 \right]} \; . \label{mzs} 
\end{eqnarray}
We defined $\lambda_{L/R}$ 
\begin{eqnarray}
\lambda_L \equiv \kappa_L \sqrt{L_e\,L_q} \; ,
\label{lambdall} \\
\lambda_R \equiv \kappa_R \sqrt{R_e\,R_q} \; ,
\label{lambdarr}
\end{eqnarray}
with $R_f$ ($L_f$) being the right-handed (left-handed) coupling of the
fermion $f$ to the $Z$, defined as
\begin{eqnarray}
L_f&=& 2(T^f_3 - Q_f \sin^2{\theta_W}) \;,\\
R_f&=& -2Q_f \sin^2{\theta_W} \;,
\end{eqnarray}
where $Q_f$ is the electromagnetic charge of the fermion, $T^f_3$ is the third 
component of the isospin, and $\theta_W$ is the weak angle.

For $F=2$ vector leptoquarks, we have
\begin{eqnarray}
\left(\frac{d\hat{\sigma}}{d\cos\theta}\right)_{e^-\,q \to e^-\,q}^{V}
&=& \frac{1}{32\pi\hat{s}}\,\left[ |{\cal M}_{\gamma}|^2 + |{\cal
    M}_{Z^0}|^2 + |{\cal M}_V|^2 + 2{\rm Re}({\cal M}_{\gamma Z^0})  \right. 
\nonumber \\
&+& \left. 2{\rm Re}({\cal M}_{\gamma V}) + 2{\rm Re}({\cal M}_{Z^0
    V}) \right] \; ,
\label{dsdcosvet}
\end{eqnarray}
with
\begin{eqnarray}
|{\cal M}_V|^2 &=& \kappa^4 \frac{\hat{u}^2}{(\hat{s} - m_V^2)^2 + m_V^2
  \Gamma_V^2} \; ,
\label{mv2} \\
{\cal M}_{\gamma V} &=& 2 Q_q \kappa^2 e^2\, \frac{\hat{u}^2 (\hat{s} -
  m_V^2)}{\hat{t} \left[ (\hat{s} - m_V^2)^2 + m_V^2 \Gamma_V^2 \right]}
\; ,
\label{mgv} \\ 
2{\rm Re}({\cal M}_{Z^0 V}) &=& -\frac{4 \lambda^{\prime 2}_{L/R} 
G_F m_{Z^0}^2}{\sqrt{2}} 
\frac{\hat{u}^2 (\hat{s} - m_V^2)}{(\hat{t} -
  m_{Z^0}^2) \left[ (\hat{s} - m_V^2)^2 + m_V^2 \Gamma_V^2 \right]} \; ,
\label{mzv}
\end{eqnarray}
where
\begin{eqnarray}
\lambda^{\prime}_L \equiv \kappa_L \sqrt{L_e\,R_q} \; ,
\label{lambdaprimel} \\
\lambda^{\prime}_R \equiv \kappa_R \sqrt{R_e\,L_q} \; .
\label{lambdaprimer}
\end{eqnarray}

The differential cross section of the process $e^-\,\bar{q}\to e^-\,\bar{q}$,
taking into account the contribution of a $F=2$ scalar leptoquark is
\begin{eqnarray}
\left(\frac{d\hat{\sigma}}{d\cos\theta}\right)_{e^-\,\bar{q} \to 
e^-\,\bar{q}}^{S} &=& \frac{1}{32\pi\hat{s}}\,\left[ |{\cal 
M}_{\gamma}|^2 + |{\cal M}^{\prime}_{Z^0}|^2 + |{\cal M}^{\prime}_S|^2
+ 2{\rm Re}({\cal M}^{\prime}_{\gamma Z^0}) \right.
\nonumber \\
&+& \left. 2{\rm Re}({\cal M}^{\prime}_{\gamma S}) + 
2{\rm Re}({\cal M}^{\prime}_{Z^0 S}) \right] \; ,
\label{sigmas}
\end{eqnarray}
with ${\cal M}^{\prime}$ given by (\ref{mz2}) and (\ref{mzs}) just switching
$\hat{s} \leftrightarrow \hat{u}$ and $|{\cal M}_{\gamma}|^2$ given by
(\ref{mg2}). The cross section for this process including vector leptoquarks
is
\begin{eqnarray}
\left(\frac{d\hat{\sigma}}{d\cos\theta}\right)_{e^-\,\bar{q} 
\to e^-\,\bar{q}}^{V}
&=& \frac{1}{32\pi\hat{s}}\,\left[ |{\cal M}_{\gamma}|^2 + |{\cal 
M}^{\prime}_{Z^0}|^2 + |{\cal M}^{\prime\prime}_V|^2 + 
2{\rm Re}({\cal M}^{\prime}_{\gamma Z^0})  \right. 
\nonumber \\
&+& \left. 2{\rm Re}({\cal M}^{\prime\prime}_{\gamma V}) + 
2{\rm Re}({\cal M}^{\prime\prime}_{Z^0 V}) 
\right] \; ,
\label{sigmav}
\end{eqnarray}
where ${\cal M}^{\prime\prime}_{\gamma(Z)V}$ are given by (\ref{mv2}) and
(\ref{mzv}) with the change $\hat{s} \leftrightarrow \hat{u}$ and the other
terms remain unchanged.

Now, we show the non--polarized differential cross sections for $F=0$ scalar
and vector leptoquarks. The cross section for the process $e^+\,q\to e^+\,q$
including a scalar leptoquark is
\begin{eqnarray}
\left(\frac{d\hat{\sigma}}{d\cos\theta}\right)_{e^+\,q \to 
e^+\,q}^{R} &=& \frac{1}{32\pi\hat{s}}\,\left[ |{\cal 
M}_{\gamma}|^2 + |{\cal M}_{Z^0}|^2 + |{\cal M}_R|^2
+ 2{\rm Re}({\cal M}_{\gamma Z^0})  \right. 
\nonumber \\
&+& \left. 2{\rm Re}({\cal M}_{\gamma R}) + 
2{\rm Re}({\cal M}_{Z^0 R}) \right] \; ,
\label{sigmar}
\end{eqnarray}
with $|{\cal M}_{\gamma}|^2$, $|{\cal M}_{Z^0}|^2$, and $2{\rm Re}({\cal
  M}_{\gamma Z^0})$ given by (\ref{mg2}), (\ref{mz2}), and (\ref{mgz})
respectively with the exchange $\hat{s} \leftrightarrow \hat{u}$.  The
remaining contributions are
\begin{eqnarray}
|{\cal M}_R|^2 &=& \frac{\kappa^4}{4}\frac{\hat{s}^2}{(\hat{s} -
  m_R^2)^2 + m_R^2\,\Gamma_R^2} \; ,
\label{mr2} \\
2{\rm Re}({\cal M}_{\gamma R}) &=& -Q_q e^2 \kappa^2 \frac{\hat{s}^2
  (\hat{s} - m_R^2)}{\hat{t}\left[ (\hat{s} - m_R^2)^2 + m_R^2\,\Gamma_R^2
  \right]} \; ,
\label{mgr} \\
2{\rm Re}({\cal M}_{Z^0 R}) &=& \frac{2 G_F m_{Z^0}^2 \eta_{L/R}^2}{\sqrt{2}}
\frac{\hat{s}^2 (\hat{s} - m_R^2)}{(\hat{t} - m_{Z^0}^2) \left[
    (\hat{s} - m_R^2)^2 + m_R^2\,\Gamma_R^2 \right]} \; , 
\label{mzr}
\end{eqnarray}
with
\begin{eqnarray}
\eta_L = \kappa_L \sqrt{L_e\,R_q} \; ,
 \label{etal} \\
\eta_R = \kappa_R \sqrt{R_e\,L_q} \; . 
\label{etar}
\end{eqnarray}

The cross section of the process $e^+\,q\to e^+\,q$ including $F=0$ vector
leptoquarks is
\begin{eqnarray}
\left(\frac{d\hat{\sigma}}{d\cos\theta}\right)_{e^+\,q 
\to e^+\,q}^{U}
&=& \frac{1}{32\pi\hat{s}}\,\left[ |{\cal M}_{\gamma}|^2 + |{\cal 
M}_{Z^0}|^2 + |{\cal M}_U|^2 + 
2{\rm Re}({\cal M}_{\gamma Z^0})  \right. 
\nonumber \\
&+& \left. 2{\rm Re}({\cal M}_{\gamma U}) + 
2{\rm Re}({\cal M}_{Z^0 U}) 
\right] \; ,
\label{sigmau}
\end{eqnarray}
where
\begin{eqnarray}
|{\cal M}_U|^2 &=& \kappa^4 \frac{\hat{u}^2}{(\hat{s} - m_U^2)^2 +
  m_U^2\, \Gamma_U^2} \; , 
\label{mu2} \\
2{\rm Re}({\cal M}_{\gamma U}) &=& -\frac{2 Q_q e^2 \kappa^2}{3}
\frac{\hat{u}^2 (\hat{s} - m_U^2)}{\hat{t} \left[ (\hat{s}
    - m_U^2)^2 + m_U^2\,\Gamma_U^2 \right]} \; , 
\label{mgu} \\
2{\rm Re}({\cal M}_{Z^0 U}) &=& \frac{4 G_F m_{Z^0}^2 \eta_{L/R}^{\prime
    2}}{\sqrt{2}} \frac{\hat{u}^2 (\hat{s} - m_U^2)}{(\hat{t} -
  m_{Z^0}^2) \left[ (\hat{s} - m_U^2)^2 + m_U^2\,\Gamma_U^2 \right]}
\; . \label{mzu}
\end{eqnarray}
We introduced the definitions
\begin{eqnarray}
\eta_L^{\prime} = \kappa_L \sqrt{L_e\,L_q} \; ,
\label{etaprimel}\\
\eta_R^{\prime} = \kappa_R \sqrt{R_e\, R_q} \; .
\label{etaprimer}
\end{eqnarray}

Finally, the cross section of the process $e^+\,\bar{q}\to e^+\,\bar{q}$,
taking into account $F=0$ scalar leptoquarks is
\begin{eqnarray}
\left(\frac{d\hat{\sigma}}{d\cos\theta}\right)_{e^+\,\bar{q} \to 
e^+\,\bar{q}}^{R} &=& \frac{1}{32\pi\hat{s}}\,\left[ |{\cal 
M}_{\gamma}|^2 + |{\cal M}^{\prime}_{Z^0}|^2 + |{\cal M}^{\prime}_R|^2
+ 2{\rm Re}({\cal M}^{\prime}_{\gamma Z^0})  \right. 
\nonumber \\
&+& \left. 2{\rm Re}({\cal M}^{\prime}_{\gamma R}) + 
2{\rm Re}({\cal M}^{\prime}_{Z^0 R}) \right] \; ,
\label{sigmar2}
\end{eqnarray}
with $|{\cal M}_{\gamma}|^2$, $|{\cal M}^{\prime}_{Z^0}|^2$, and $2{\rm
  Re}({\cal M}^{\prime}_{\gamma Z^0})$ the same as in (\ref{mg2}), (\ref{mz2})
and (\ref{mgz}) respectively, and $|{\cal M}^{\prime}_R|^2$, $2{\rm Re}({\cal
  M}^{\prime}_{\gamma R})$, and $2{\rm Re} ({\cal M}^{\prime}_{Z^0 R})$ given
by (\ref{mr2}), (\ref{mgr}) by (\ref{mzr}) respectively with the change
$\hat{s} \leftrightarrow \hat{u}$.

For the vector leptoquarks, the cross section of this last process is
\begin{eqnarray}
\left(\frac{d\hat{\sigma}}{d\cos\theta}\right)_{e^+\,\bar{q} 
\to e^+\,\bar{q}}^{U}
&=& \frac{1}{32\pi\hat{s}}\,\left[ |{\cal M}_{\gamma}|^2 + |{\cal 
M}^{\prime}_{Z^0}|^2 + |{\cal M}^{\prime\prime}_U|^2 + 
2{\rm Re}({\cal M}^{\prime}_{\gamma Z^0})  \right. 
\nonumber \\
&+& \left. 2{\rm Re}({\cal M}^{\prime\prime}_{\gamma U}) + 
2{\rm Re}({\cal M}^{\prime\prime}_{Z^0 U}) 
\right] \; ,
\label{sigmau2}
\end{eqnarray}
with ${\cal M}^{\prime\prime}$ given by (\ref{mu2}) to (\ref{mzu}) switching
$\hat{s} \leftrightarrow \hat{u}$ and the other terms are the same as
presented in (\ref{sigmar2}).




\begin{table}
\begin{center}
\begin{tabular}{|c|r|c|c|}
  $LQ$ & $Q_{em}$ & decay channels & Coupling ($\kappa_{L/R}$) 
\\ \hline
  \footnotesize{$S_1$}& $-1/3$   & $d\;\nu_e$  & $-g_{1L}$
\\
       & $-1/3$   & $u\;e^-$    & $g_{1L}$~;~$g_{1R}$ 
\\ 
\hline 
  \footnotesize{$\tilde{S}_1$} & $-4/3$ & $d\;e^-$ & $\tilde{g}_{1R}$\\
\hline 
  \footnotesize{$S^+_{3}$} & $2/3$& $u\;\nu_e$  & $\sqrt{2}g_{3L}$ \\
  \footnotesize{$S^-_{3}$} & $-4/3$ & $d\;e^-$  & $\sqrt{2}g_{3L}$ \\
  \footnotesize{$S^0_{3}$} & $-1/3$ & $d\;\nu_e$& $-g_{3L}$ \\
           & $-1/3$ & $u\;e^-$  & $-g_{3L}$ \\ 
\hline 
           & $-4/3$ & $d\;e^-$  & $g_{2L}$~;~$-g_{2R}$ \\
  \footnotesize{$V_{2\mu}$}&$-1/3$ & $u\;e^-$  & $g_{2R}$ \\
           & $-1/3$ & $d\;\nu_e$& $-g_{2L}$ 
\\ 
\hline 
 \footnotesize{$\tilde{V}_{2\mu}$}&$-1/3$&$u\;e^-$&$\tilde{g}_{2L}$ \\
           & $2/3$  & $u\;\nu_e$& $-\tilde{g}_{2L}$ \\ \hline
           & $5/3$  & $u\;e^+$  & $h_{2L}$~;~$h_{2R}$ \\
  \footnotesize{$R_2$}    & $2/3$  & $u\;\bar{\nu}_e$&$h_{2L}$ \\
           & $2/3$  & $d\;e^+$  & $-h_{2R}$ 
\\ 
\hline 
  \footnotesize{$\tilde{R}_2$}&$-1/3$&$d\;\bar{\nu}_e$&$\tilde{h}_{2L}$ \\
           & $2/3$  & $d\;e^+$ & $\tilde{h}_{2L}$ 
\\ 
\hline 
 \footnotesize{$U_{1\mu}$}&$2/3$  & $u\;\bar{\nu}_e$ & $h_{1L}$ \\
           & $2/3$  & $d\;e^+$  & $h_{1L}$~;~$h_{1R}$ 
\\ 
\hline 
  \footnotesize{$\tilde{U}_{1\mu}$}&$5/3$&$u\;e^+$&$\tilde{h}_{1R}$ \\
\hline 
  \footnotesize{$U^{+\mu}_{3}$} & $-1/3$ & $d\;\bar{\nu}_e$ &
 $\sqrt{2}h_{3L}$ \\ 
  \footnotesize{$U^{-\mu}_{3}$} & $5/3$ & $u\;e^+$ & $\sqrt{2}h_{2L}$ \\
  \footnotesize{$U^{0\mu}_{3}$} & $2/3$ & $u\;\bar{\nu}_e$ & $h_{3L}$ \\
           & $2/3$ & $d\;e^+$ & $-h_{3L}$ \\ 
\end{tabular}
\caption{Scalar and vector leptoquarks that can be observed through their 
  decays into a $e^\pm$ and a jet and the correspondent decay channels. For
  simplicity we introduced the left- and right-handed leptoquarks in the same
  entry.}
\label{t:cor}
\end{center}
\end{table}


\begin{table}
\begin{center}
\begin{tabular}{||l|c|c|c|c|c|c|c|c||}
$M_{\text{lq}}$ (GeV) & 500& 750& 1000& 1250& 1500& 2000& 2500&  3000 \\
$p_T^{\text{min}}$ (GeV)& 200& 300&  400&  500&  600&  800& 1000&  1000 \\
$\Delta M$ (GeV)        &  40 &  40 & 40 & 50 & 50 & 50 & 60 &  60 \\
$p_T^{\text{miss}}$ (GeV)  & 30 & 30 & 30 & 60 & 60 & 60 & 80 &  80  
\end{tabular}
\medskip\medskip
\caption{Values of the cuts $p_T^{\text{min}}$, 
  $\Delta M$, and $p_T^{\text{miss}}$ for several leptoquark masses.}
\label{bins}
\end{center}
\end{table}

\begin{table}
\begin{center}
\begin{tabular}{||c|c|c||c|c||}
 & \multicolumn{2}{l||}{$e^-$ + jet} &\multicolumn{2}{l||}{$e^+$ + 
jet}\\ \hline 
 & C1 + C2 & C1--C6 & C1 + C2 & C1--C6
 \\ \hline
$\tilde{S}_1$ &16.3 &2.30 &9.98 &0.576 \\ \hline
$S_{1L}$ &16.9 &2.45 &9.02 &0.299 \\ \hline
$S_{1R}$ &26.0 &4.95 &10.0 &0.579 \\ \hline
$S_{3}$ &32.9 &6.86 &12.9 &1.37 \\ \hline
$V^\mu_{2L}$ &24.6 &3.85 &12.0 &0.953 \\ \hline
$V^\mu_{2R}$ &60.6 &12.1 &16.0 &1.88 \\ \hline
$\tilde{V}^\mu_{2}$ &44.1 &8.32 &12.1 &0.957 \\ \hline
$\tilde{R}_2$ &10.1 &0.595 &16.2 &2.28 \\ \hline
$R_{2L}$ &10.1 &0.594 &25.8 &4.93 \\ \hline
$R_{2R}$ &12.1 &1.14 &34.0 &7.18 \\ \hline
$U^\mu_{3}$ &17.9 &2.32 &87.0 &18.3 \\ \hline
$U^\mu_{1L}$ &10.1 &0.505 &16.2 &1.91 \\ \hline
$U^\mu_{1R}$ &12.2 &0.971 &24.5 &3.83 \\ \hline
$\tilde{U}^\mu_{1}$ &12.1 &0.965 &43.9 &8.30 \\ \hline
$\sigma_{bg}$ &8.14 &0.0451 &8.00 &0.0285 \\ \hline
$\sigma_{W}$ &68.3 &1.05 &68.5 &0.928 \\ \hline
$\sigma_{Z}$ &33.9 &0.050 &31.7 &0.013 
\end{tabular}
\medskip\medskip
\caption{Total cross section in fb for the signals and backgrounds for 
  all leptoquark multiplets after and before applying the cuts C3--C6. We
  assumed $M_{\text{lq}} = 1$ TeV and $\kappa = 0.3$ and smeared all the final
  state momenta.}
\label{uncut:signal}
\end{center}
\end{table}


\begin{table}
\begin{center}
\begin{tabular}{||c|c||}
Leptoquark & $M_{\text{lq}}$ (TeV) \\ \hline
$S_{1L}$        &       2.04 \\
$S_{1R}$        &       2.40 \\
$\tilde{S}_1$   &       1.92 \\
$S_{3BC}$       &       2.47 \\
$V_{2L}$        &       2.15 \\
$V_{2R}$        &       2.92 \\
$\tilde{V}_2$   &       2.73 \\
$R_{2L}$        &       2.40 \\
$R_{2R}$        &       2.56 \\
$\tilde{R}_2$   &       1.92 \\
$U_{1L}$        &       1.85 \\
$U_{1R}$        &       2.14 \\
$\tilde{U}_1$   &       2.73 \\
$U_{3BC}$       &       3.21 
\end{tabular}
\medskip\medskip
\caption{Attainable limits for the different leptoquark multiplets
  at 99.7\% CL (3$\sigma$), assuming $\kappa = 0.3$ and an integrated
  luminosity of 100 fb$^{-1}$.}
\label{tab:lim}
\end{center}
\end{table}


\newpage

\newpage

\newpage

\newpage

\newpage

\newpage

\end{document}